\documentclass[aps,prd,amssymb,twocolumn,showpacs,letterpaper,superscriptaddress,10pt,floatfix]{revtex4-2}

\usepackage{amsmath}
\usepackage{color}
\usepackage[dvipsnames]{xcolor}
\usepackage{graphics}
\usepackage{epsfig}
\usepackage{subfigure}
\DeclareUnicodeCharacter{03B3}{$\gamma$}
\usepackage{hyperref} 

\begin{document}

\title{Defect Formation in NaI Crystals: A Novel Pathway to Dark Matter Detection}

\author{G.~Angloher}
\affiliation{Max-Planck-Institut f\"ur Physik, 85748 Garching, Germany}

\author{M.~R.~Bharadwaj}
\affiliation{Max-Planck-Institut f\"ur Physik, 85748 Garching, Germany}

\author{A.~B\"ohmer}
\affiliation{Marietta-Blau-Institut für Teilchenphysik der Österreichischen Akademie der Wissenschaften, 1010 Wien, Austria}
\affiliation{Atominstitut, Technische Universit\"at Wien, 1020 Wien, Austria}

\author{M.~Cababie}
\affiliation{Marietta-Blau-Institut für Teilchenphysik der Österreichischen Akademie der Wissenschaften, 1010 Wien, Austria}
\affiliation{Atominstitut, Technische Universit\"at Wien, 1020 Wien, Austria}

\author{I.~Colantoni}
\affiliation{Consiglio Nazionale delle Ricerche, Istituto di Nanotecnologia, 00185 Roma, Italy}
\affiliation{INFN - Sezione di Roma, 00185 Roma, Italy}

\author{I.~Dafinei}
\affiliation{Gran Sasso Science Institute, 67100 L'Aquila, Italy}
\affiliation{INFN - Sezione di Roma, 00185 Roma, Italy}

\author{N.~Di~Marco}
\affiliation{Gran Sasso Science Institute, 67100 L'Aquila, Italy}
\affiliation{INFN - Laboratori Nazionali del Gran Sasso, 67100 Assergi, Italy}

\author{C.~Dittmar}
\affiliation{Max-Planck-Institut f\"ur Physik, 85748 Garching, Germany}

\author{F.~Ferella}
\affiliation{Dipartimento di Scienze Fisiche e Chimiche, Universit\`a degli Studi dell'Aquila, 67100 L'Aquila, Italy}
\affiliation{INFN - Laboratori Nazionali del Gran Sasso, 67100 Assergi, Italy}

\author{F.~Ferroni}
\affiliation{INFN - Sezione di Roma, 00185 Roma, Italy}
\affiliation{Gran Sasso Science Institute, 67100 L'Aquila, Italy}

\author{S.~Fichtinger}
\affiliation{Marietta-Blau-Institut für Teilchenphysik der Österreichischen Akademie der Wissenschaften, 1010 Wien, Austria}

\author{A.~Filipponi}
\affiliation{Dipartimento di Scienze Fisiche e Chimiche, Universit\`a degli Studi dell'Aquila, 67100 L'Aquila, Italy}
\affiliation{INFN - Laboratori Nazionali del Gran Sasso, 67100 Assergi, Italy}

\author{T.~Frank}
\affiliation{Max-Planck-Institut f\"ur Physik, 85748 Garching, Germany}

\author{M.~Friedl}
\affiliation{Marietta-Blau-Institut für Teilchenphysik der Österreichischen Akademie der Wissenschaften, 1010 Wien, Austria}

\author{D.~Fuchs}
\affiliation{Marietta-Blau-Institut für Teilchenphysik der Österreichischen Akademie der Wissenschaften, 1010 Wien, Austria}
\affiliation{Atominstitut, Technische Universit\"at Wien, 1020 Wien, Austria}

\author{L.~Gai}
\affiliation{State Key Laboratory of Functional Crystals and Devices, Shanghai Institute of Ceramics, Chinese Academy of Sciences, 201899 Shanghai, China}

\author{M.~Gapp}
\affiliation{Max-Planck-Institut f\"ur Physik, 85748 Garching, Germany}

\author{M.~Heikinheimo}
\affiliation{Helsinki Institute of Physics, 00014 University of Helsinki, Finland}

\author{M.~N.~Hughes}
\affiliation{Max-Planck-Institut f\"ur Physik, 85748 Garching, Germany}

\author{K.~Huitu}
\affiliation{Helsinki Institute of Physics, 00014 University of Helsinki, Finland}

\author{M.~Kellermann}
\affiliation{Max-Planck-Institut f\"ur Physik, 85748 Garching, Germany}

\author{R.~Maji}
\affiliation{Marietta-Blau-Institut für Teilchenphysik der Österreichischen Akademie der Wissenschaften, 1010 Wien, Austria}
\affiliation{Atominstitut, Technische Universit\"at Wien, 1020 Wien, Austria}

\author{M.~Mancuso}
\affiliation{Max-Planck-Institut f\"ur Physik, 85748 Garching, Germany}

\author{L.~Pagnanini}
\affiliation{Gran Sasso Science Institute, 67100 L'Aquila, Italy}
\affiliation{INFN - Laboratori Nazionali del Gran Sasso, 67100 Assergi, Italy}

\author{F.~Petricca}
\affiliation{Max-Planck-Institut f\"ur Physik, 85748 Garching, Germany}

\author{S.~Pirro}
\affiliation{INFN - Laboratori Nazionali del Gran Sasso, 67100 Assergi, Italy}

\author{F.~Pr\"obst}
\affiliation{Max-Planck-Institut f\"ur Physik, 85748 Garching, Germany}

\author{G.~Profeta}
\affiliation{Dipartimento di Scienze Fisiche e Chimiche, Universit\`a degli Studi dell'Aquila, 67100 L'Aquila, Italy}
\affiliation{INFN - Laboratori Nazionali del Gran Sasso, 67100 Assergi, Italy}

\author{A.~Puiu}
\affiliation{INFN - Laboratori Nazionali del Gran Sasso, 67100 Assergi, Italy}

\author{F.~Reindl}
\affiliation{Marietta-Blau-Institut für Teilchenphysik der Österreichischen Akademie der Wissenschaften, 1010 Wien, Austria}
\affiliation{Atominstitut, Technische Universit\"at Wien, 1020 Wien, Austria}

\author{K.~Sch\"affner}
\affiliation{Max-Planck-Institut f\"ur Physik, 85748 Garching, Germany}

\author{J.~Schieck}
\affiliation{Marietta-Blau-Institut für Teilchenphysik der Österreichischen Akademie der Wissenschaften, 1010 Wien, Austria}
\affiliation{Atominstitut, Technische Universit\"at Wien, 1020 Wien, Austria}

\author{P.~Schreiner}
\affiliation{Marietta-Blau-Institut für Teilchenphysik der Österreichischen Akademie der Wissenschaften, 1010 Wien, Austria}
\affiliation{Atominstitut, Technische Universit\"at Wien, 1020 Wien, Austria}

\author{C.~Schwertner}
\affiliation{Marietta-Blau-Institut für Teilchenphysik der Österreichischen Akademie der Wissenschaften, 1010 Wien, Austria}
\affiliation{Atominstitut, Technische Universit\"at Wien, 1020 Wien, Austria}

\author{P.~Settembri}
\thanks{Corresponding author: \href{mailto: paolo.settembri@cern.ch}{paolo.settembri@cern.ch}}
\affiliation{Dipartimento di Scienze Fisiche e Chimiche, Universit\`a degli Studi dell'Aquila, 67100 L'Aquila, Italy}
\affiliation{INFN - Laboratori Nazionali del Gran Sasso, 67100 Assergi, Italy}

\author{K.~Shera}
\affiliation{Max-Planck-Institut f\"ur Physik, 85748 Garching, Germany}

\author{M.~Stahlberg}
\affiliation{Max-Planck-Institut f\"ur Physik, 85748 Garching, Germany}

\author{A.~Stendahl}
\affiliation{Helsinki Institute of Physics, 00014 University of Helsinki, Finland}

\author{M.~Stukel}
\affiliation{SNOLAB, P3Y 1N2 Lively, Canada}
\affiliation{INFN - Laboratori Nazionali del Gran Sasso, 67100 Assergi, Italy}

\author{C.~Tresca}
\affiliation{CNR-SPIN c/o Dipartimento di Scienze Fisiche e Chimiche, Universit\`a degli Studi dell'Aquila, 67100 L'Aquila, Italy}
\affiliation{INFN - Laboratori Nazionali del Gran Sasso, 67100 Assergi, Italy}

\author{S.~Yue}
\affiliation{SICCAS - Shanghai Institute of Ceramics, Shanghai, P.R.China 200050}

\author{V.~Zema}
\affiliation{Max-Planck-Institut f\"ur Physik, 85748 Garching, Germany}
\affiliation{Marietta-Blau-Institut für Teilchenphysik der Österreichischen Akademie der Wissenschaften, 1010 Wien, Austria}

\author{Y.~Zhu}
\affiliation{SICCAS - Shanghai Institute of Ceramics, Shanghai, P.R.China 200050}

\author{N.~Zimmermann}
\affiliation{Helsinki Institute of Physics, 00014 University of Helsinki, Finland}

\collaboration{The COSINUS Collaboration}
\noaffiliation

\author{M.~Di Giambattista}
\affiliation{Dipartimento di Scienze Fisiche e Chimiche, Universit\`a degli Studi dell'Aquila, 67100 L'Aquila, Italy}

\author{F.~Giannessi}
\affiliation{Dipartimento di Scienze Fisiche e Chimiche, Universit\`a degli Studi dell'Aquila, 67100 L'Aquila, Italy}

\author{R.~Rollo}
\affiliation{Dipartimento di Scienze Fisiche e Chimiche, Universit\`a degli Studi dell'Aquila, 67100 L'Aquila, Italy}

\begin{abstract}
{
Sodium iodide (NaI) is a widely used scintillator in direct dark matter searches. In particular, NaI-based cryogenic scintillating calorimeters have emerged as promising candidates, like in the \mbox{COSINUS} experiment, for testing the annually modulating signal reported by \mbox{DAMA/LIBRA}. In this study, we investigate defect formation within NaI crystals and its impact on the dark matter detection signal. Using molecular dynamics simulations and density functional theory techniques, we simulate a DM particle collision on an NaI crystal, focusing on the possible defects formation and their structural and electronic properties. Our analysis includes a detailed study of the electronic states associated with the interstitial atoms and vacancies, the energetic cost of defect formation, and the anisotropic threshold displacement energy. Finally, we highlight the potential to exploit dark matter-induced defects as a novel detection channel, enabled by the introduction of new states within the electronic band gap.
}
\end{abstract}

\maketitle

\section{Introduction}
The search for evidence of dark matter (DM), namely a form of matter of which a large part of our Universe is composed, has been one of the main research topics in the last 50 years~\cite{DarkMatter1,DarkMatter2,DarkMatter3}. 
Some cosmological and astrophysical models suggest that dark matter has a particle-like nature, weakly interacting with ordinary matter (WIMP, Weakly Interacting Massive Particle)~\cite{WIMP}, making possible its direct detection in terrestrial laboratories.
Actually, many Earth-located experiments are designed to detect DM particles interacting with ordinary matter (in a solid, liquid or gaseous state), like: \mbox{ANAIS–112}~\cite{ANAIS}, \mbox{COSINE-100}~\cite{COSINE100}, \mbox{CRESST}~\cite{CRESST}, \mbox{DAMA/LIBRA}~\cite{DAMA/LIBRA}, \mbox{DAMIC}~\cite{DAMIC}, \mbox{DarkSide-50}~\cite{Darkside50}, \mbox{DM-Ice}~\cite{DMICE}, \mbox{KIMS}~\cite{KIMS}, \mbox{LUX}~\cite{LUX}, \mbox{SABRE}~\cite{SABRE}, \mbox{SuperCDMS}~\cite{supercdms}, \mbox{XENONnT}~\cite{xenonnt}, and others~\cite{EDELWEISS,LUX-ZEPLIN,deap,CoGeNT}.
All these experiments differ in the material used as a target, and the output channel used to read the DM event's signal.
Despite significant efforts by the scientific community, no clear findings have yet been made regarding the direct detection of DM.
Among the most discussed and relevant results in the field is the annual modulation signal observed by \mbox{DAMA/LIBRA}, achieved with a statistical significance of 13.7$\sigma$, using NaI(Tl) crystals as room temperature scintillating detectors~\cite{DAMA2,bernabei-2022}. 
Other direct detection experiments have set exclusion limits that conflict with the \mbox{DAMA/LIBRA} results. However, excluding a DM explanation for the \mbox{DAMA/LIBRA} signal is challenging, as comparisons between experiments depend on assumptions about both the astrophysical distribution and particle physics properties of DM. These conclusions can be highly sensitive to these assumptions. An independent test of the \mbox{DAMA/LIBRA} result using NaI-based detectors would minimize the need for such assumptions~\cite{COSINUS4_model}. Current efforts in this direction include the \mbox{ANAIS-112}~\cite{ANAIS} and \mbox{COSINE-100}~\cite{COSINE100} experiments, both employing Thallium-doped NaI(Tl) crystals to independently test the DAMA/LIBRA modulation.
\\
\indent The \mbox{COSINUS} experiment~\cite{COSINUS1,COSINUS2,COSINUS3,cosinus2025} aims at providing a model-independent comparison, by using an NaI target as a cryogenic scintillating calorimeter, providing two detection channels.
Assuming a DM particle releases energy through a collision in the NaI system, it will cause the emission of scintillation light, due to electronic excitations, and will produce phonons, due to nuclear recoils. This gives rise to the two possible detection channels: capturing the scintillation light and measuring the phonons at the target surface. 
Having multiple detection channels enables event-by-event discrimination of interaction types, aiding in distinguishing signal events from background.\\
\indent In addition to the light and thermal channels, detectors can also be sensitive to a charge signal, which could serve as an additional channel to strengthen experimental results.
Recent studies have explored the potential for a charge signal channel arising from defect formation induced by low-mass DM in solid state detectors~\cite{Kadribasic_2018}.
These defects, created by the collision of DM particles with the target material, can also alter the phonon measurements by redirecting energy into their formation~\cite{Sassi1,Sassi2}. The \mbox{CRESST} collaboration recently reported a possible energy-scale shift and discussed defect formation in the target crystal as a potential explanation~\cite{CRESSTdefect}.\\
\indent In this work we investigate defect formation in NaI crystals induced by DM collisions, focusing on their impact across different detection channels. 
Combining molecular dynamics (MD) simulations with density functional theory (DFT) calculations, we analyze the electronic configuration of the system in the presence of defects and characterize their properties.
By calculating the DM-nuclei scattering rate, we identify the range of DM masses where defect formation becomes significant.
Finally, we discuss the influence of these defects on the detection rate, highlighting the potential for a novel charge signal in NaI-based detectors.

\section{Defects Formation}
\subsection{Computational details}
Molecular dynamics is a simulation technique that follows the motion of atoms in a material by numerically integrating Newton’s equations \cite{ComputerSimulationofLiquids,UnderstandingMolecularSimulation}. Forces acting on each atom are derived from interatomic potentials that approximate the underlying quantum-mechanical interactions \cite{InteratomicForcesinCondensedMatter}. While electronic degrees of freedom are not treated explicitly, MD captures the classical atomic trajectories that govern the creation and evolution of lattice defects on femtoseconds–picoseconds timescales.\\
Lattice defects are local disruptions of the regular atomic arrangement in a solid, that can be created by thermal fluctuations or by energetic events, like recoiling atoms, and influence the local structure and properties of the material \cite{crawford-1972}. Examples of such defects are vacancies (an empty lattice site) and interstitial atoms (an extra atom occupying a position in between lattice sites).
\indent We performed molecular dynamics simulations using the \mbox{LAMMPS} code~\cite{LAMMPS}.
We adopted a 6$\times$6$\times$6 unit cells configuration, with periodic boundary conditions applied on all directions.
The thermalization was performed by a combination of constant number-volume-energy (NVE) and constant number-volume-temperature (NVT) simulations, keeping the temperature at $\sim$ 1~K, until reaching the equilibrium state (after $\sim$ 2~ps).
We modeled the DM particle collision by assigning an energy $E$ and a momentum $\Vec{p}$ to an atom, called the PKA (Primary Knock-on Atom).
Then to emulate the effect of a thermal bath acting on the system, allowing the relaxation of the released energy, we divided our initial configuration in two regions: an outer region 1 unit cell thick in all directions, that performed an NVT simulation at 1~K, and an inner region that performed an NVE simulation.
To correctly describe the dynamic of the atoms after the release of high energies, we used a force field potential~\cite{ZHOU20112470}. 
To integrate correctly the equations of motion, we used an adaptive time step (going from 10$^{-5}$~fs to 1~fs) in the initial part of the simulations, when high energies and velocities are involved.
Simulations were stopped after the potential energy fluctuations fell below a threshold value of 1~eV, with a typical time of $\sim$ 5~ps. This threshold value is sufficiently smaller than the formation energy of Frenkel pairs in the MD simulations.
To analyze the final configuration, and check for the formation of Frenkel pairs, which consist of a vacancy and an interstitial atom pair, we used the Wigner-Seitz defect analysis present in \mbox{OVITO}~\cite{ovito}. 
This procedure was repeated to create a map of the energy needed to form a defect, for a given direction of the PKA recoil in the solid angle, called Threshold Displacement Energy (TDE). The TDE has been evaluated sampling the full solid angle uniformly for over 180 directions. The possibility of a scattering above the threshold value not producing a defect was not considered. The system's different configuration at the time of the collision may cause these events. But since we considered a very low temperature system, we can effectively treat it as frozen, canceling this possible effect.\\
To study the electronic properties of the system in the presence and absence of Frenkel pairs, we mapped the system on a smaller 3$\times$3$\times$3 representative subset of the simulation cell. 
The defective structures, containing both the interstitial atom and the vacancy, were derived from the molecular dynamic simulations.
Density functional theory simulations were performed on such systems. 
DFT is a first-principles quantum-mechanical method used to compute the electronic structure of materials, in which instead of solving the many-electron Schr\"odinger equation directly, the problem is reformulated in terms of the electron density, allowing practical calculations of band structures, total energies, and response functions \cite{HohenbergKohn,KohnSham,Martin_2004}.
For our DFT calculations the \mbox{VASP} code~\cite{VASP1,VASP2} was used.\\
The \mbox{PBE} (Perdew-Burke-Ernzerhof) exchange correlation functional~\cite{PBE} was used, with a kinetic energy cutoff of 350~eV, and a $k$-point grid of 6$\times$6$\times$6. 
We relaxed the structure converging the total energy up to a threshold of $10^{-5}$~eV.
For the density of states calculations a 15$\times$15$\times$15 $k$-point grid was adopted.
An electronic gaussian smearing of 0.1~eV was used for all calculations.
\subsection{Molecular Dynamics and DFT results}

\begin{figure}[t]
    \centering
    \includegraphics[width=\columnwidth]{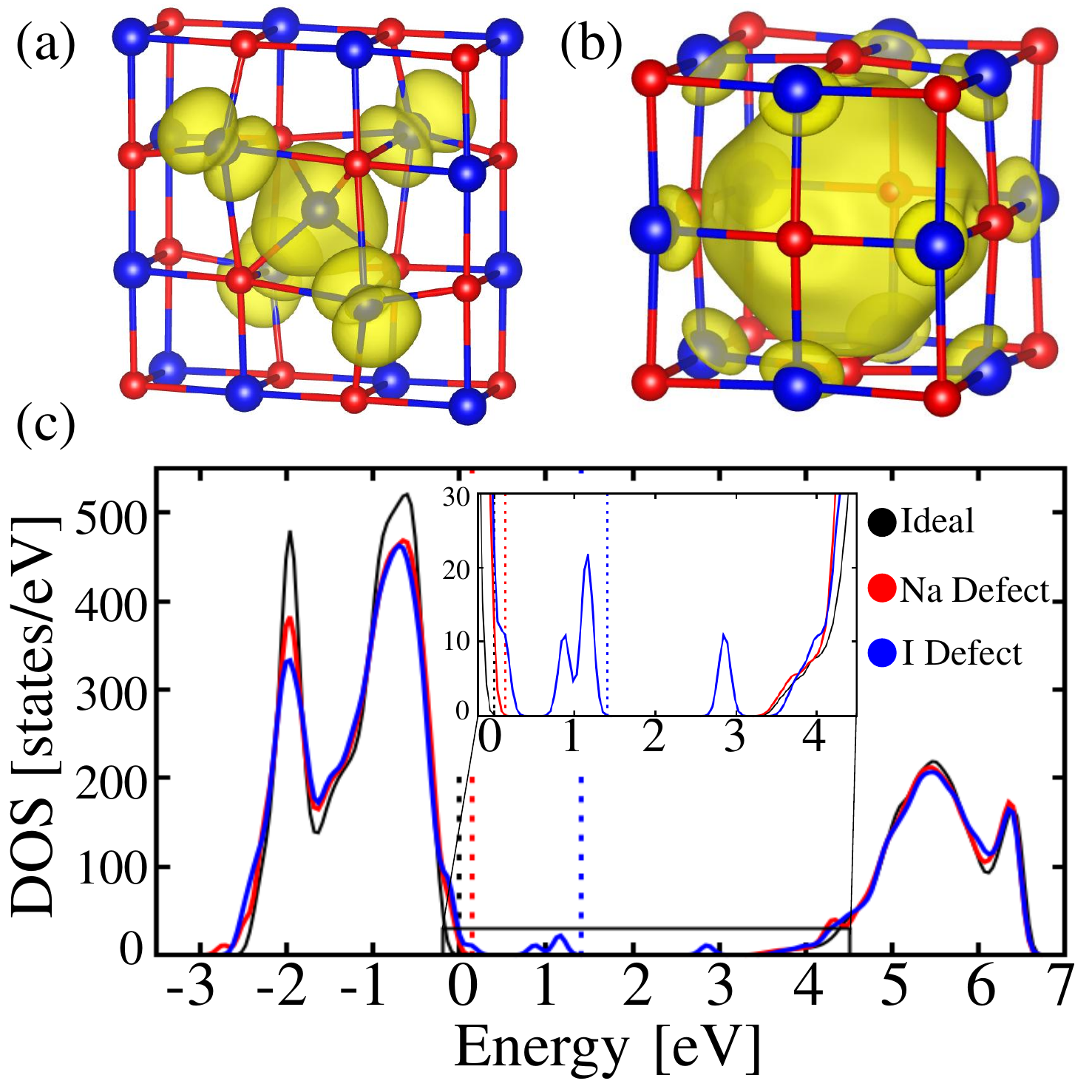}
    \caption{(a-b) Iodine interstitial atom and its corresponding vacancy. Na atoms are shown in red while I atoms are in blue. The yellow surfaces indicate the charge density associated to the states present in the electronic band gap, induced by the I defect. (c) Comparison of the electronic density of states between the ideal NaI system in absence of defects (black line), and in the presence of an interstitial Na atom (red line), and I atom (blue line). 
    The energy position of the highest occupied electronic state for each system is shown by a vertical dashed line of the corresponding color.
    An inset of the electronic band gap region is also present, showcasing how in the presence of I interstitial atoms new states emerge in the band gap.}
    \label{fig:defect-dos}
\end{figure}

The performed molecular dynamics simulations allowed the study of the required conditions for the formation of Frenkel pairs, induced by DM collisions on the NaI crystal ions.
In Fig.~\ref{fig:defect-dos}~(a-b) we report the simulated structures in the proximity of an I interstitial atom and the vacancy left from it, respectively. The equivalent case for an Na interstitial atom is reported in Appendix \hyperref[AppendixA]{A}.
It can be seen how the interstitial iodine atom, which has a far greater mass with respect to the sodium atom ($m_{\text{I}}=$ 126.90~a.m.u.; $m_{\text{Na}}=$ 22.99~a.m.u.) causes a more relevant deformation on the neighboring atoms.
In Fig.~\ref{fig:defect-dos}~(c) we report the electronic density of states (DOS), computed with DFT simulations, for an ideal (black), an Na defect (red) and an I defect (blue) system.
To better see the DOS variation due to the defect presence, the DOS of the defected supercells are shifted aligning the valence band maximum (VBM) of atoms far from the defect with the VBM of the ideal system. The major effects appear near the electronic band gap, which is shown in the inset of Fig.~\ref{fig:defect-dos}~(c).
In the iodine defect case we can see the formation of new occupied electronic states in the band gap at about 1~eV and one empty state at about 3~eV.
The real space charge densities are shown in Fig.~\ref{fig:defect-dos}~(a) and (b), respectively. The occupied states are localized around the interstitial atom and extend mainly up to its first nearest-neighbors, while the unoccupied state is localized at the vacancy position.\\
\begin{figure}
    \centering
    \includegraphics[width=\columnwidth]{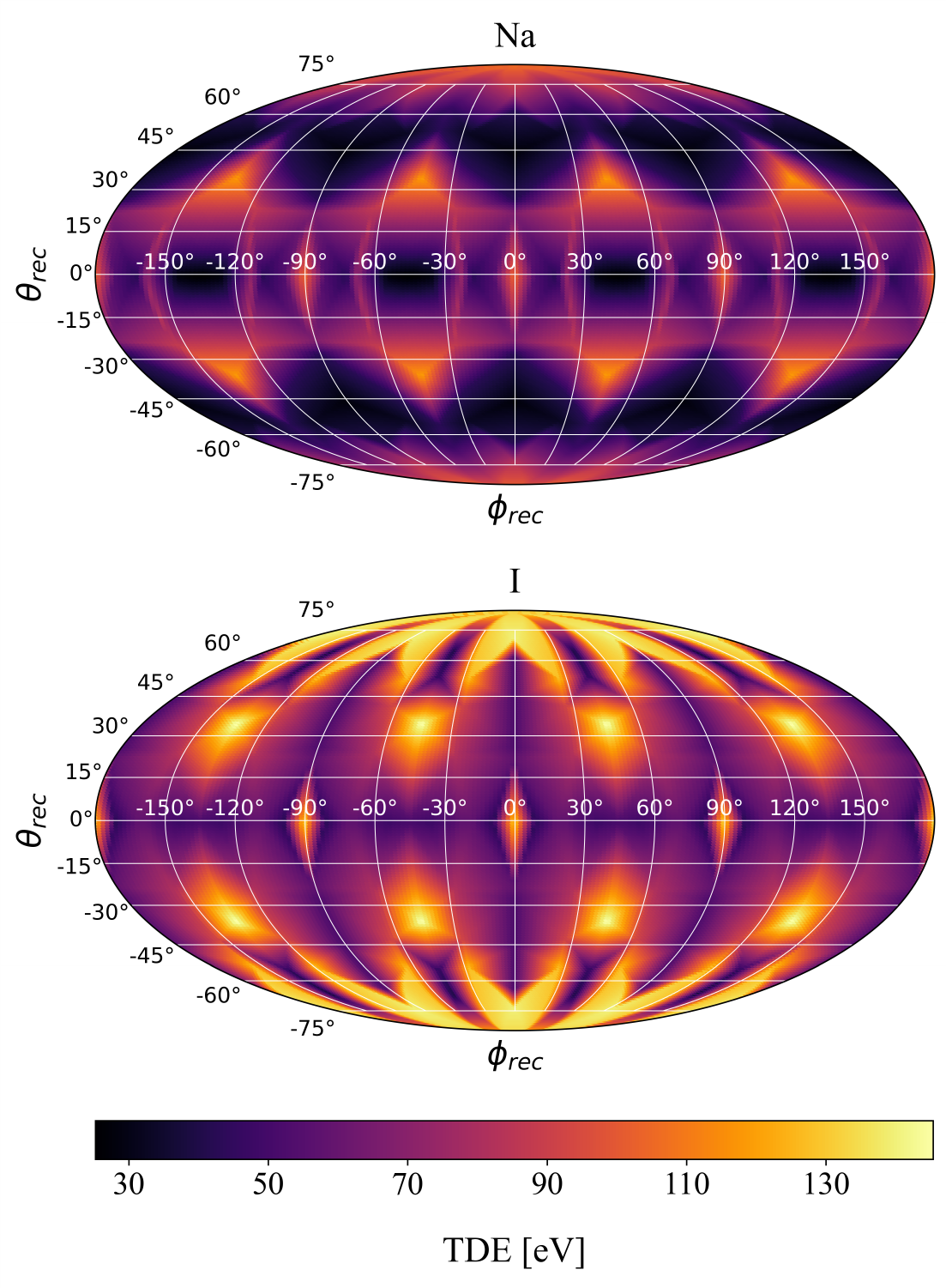}
    \caption{Mollweide projection of the TDE, in case of Na atom as the PKA (top), and the I atom (bottom). Lighter regions correspond to high TDE values, while darker ones to lower TDE values. Due to the larger mass, the TDE is larger when the PKA is an I atom.}
    \label{fig:Mollweide}
\end{figure}
By comparing the system's potential energy at the start and end of our MD simulations, we can estimate the formation energy of the Frenkel pairs, obtaining: $E_{\text{Def,Na}} \sim$ 4~eV and $E_{\text{Def,I}} \sim$ 16~eV.
To further characterize the defect formation in NaI crystals, we studied the lowest energy necessary to form a defect: the threshold displacement energy (TDE).
This quantity depends both on the recoil direction, that we identify with the polar angle $\theta_{\text{rec}}$ and the azimuthal angle $\phi_{\text{rec}}$, with respect to the [100] crystal direction, and on the atom type the DM collides with, the PKA.\\
In Fig.~\ref{fig:Mollweide} a Molleweide projection of the computed TDE is reported.
The lowest values of the TDE are in the [110] direction, with TDE$_{\text{Na}}^{[110]}=$ 24~eV and TDE$_{\text{I}}^{[110]}=$ 51~eV, while the highest TDE values are found in the [111] direction, with TDE$_{\text{Na}}^{[111]}=$ 119~eV and TDE$_{\text{I}}^{[111]}=$ 147~eV.\\
A high-energy collision can produce an interstitial atom, but the atom that ultimately occupies the interstitial site is not necessarily the PKA itself, and it may be either an ion of the same species as the PKA or of a different one. The relative probabilities of these outcomes depend on the PKA species, its recoil direction and energy, and the formation energies of the different defects. In NaI, the formation energy of an Na interstitial is significantly lower than that of an I interstitial, making Na interstitials generally more favorable.
As a result, when Na is the PKA, the first defect created is always an Na interstitial, independently of the recoil direction.
When I is the PKA, however, some recoil directions lead to a I atom occupying the interstitial position, while for other directions an interstitial Na is created, with the I PKA transferring its momentum to neighboring atoms and returning in its lattice site.
\subsection{Discussion}
The possible implications of the formation of defects are hereby discussed.\\
DM events, and in particular higher energy collisions, are expected to produce multiple defects, leading to a loss in the phonon signal, due to the defects formation energy as pointed out in Ref.~\cite{Sassi1,Sassi2}. 
As a consequence, the scintillation light energy and light yield, which quantifies the ratio between the light and phonon signal, can be strongly affected by the formation of defects, see Ref.~\cite{ScintillationDamage}. In addition, defects' concentration grows up as a function of the operational time, resulting in a degradation of the detector's light output and reduction of the energy resolution~\cite{ScintillationDamage,scintillation2,scintillation3}.\\
At the same time, the phonon signal is influenced by defect formation, as a part of the energy released during a DM collision is stored as potential energy due to structural deformations. Additionally, we find that defects induce localized vibrational modes in the system, which can also affect the phonon signal's shape with a possible increase of its tail. This is further discussed in Appendix \hyperref[AppendixA]{A}.\\
\indent Focusing on the electronic channels of detection, the appearance of defect induced in-gap states can have different consequences on the electronic channels of detection. 
They can enable electronic excitations across the band gap, from valence bands to un-occupied in-gap states or from occupied in-gap states to the conduction bands, through various mechanisms. 
For example, as the in-gap states energy depends on the interstitial atoms's position, they are modulated by the "local" temperature around the defect possibly inducing a charge transfer from in-gap states towards the conduction band, acting as an "elevator state" as proposed in Ref.~\cite{Kadribasic_2018,Elevator1,Elevator2,ElevatorDaily}.
A second possible detection mechanism associated to these states could be the measurement of defect concentration after long exposure to DM flux by means of luminescence and fluorescence techniques~\cite{Luminescence,Fluorescence}, in line with what was proposed for the paleo-detection of DM~\cite{Mineral}.
In addition, as shown in Fig.~\ref{fig:defect-dos}~(c), iodine Frenkel pair can allow previously forbidden transitions at lower energies, which become accessible only after a defect is formed by the collision, which can be exploited as photodetector.\\
It is thus interesting to investigate the effect of defects on the DM detection rate.

\section{DM signal}
To further study the effect of defects on the DM signal we must introduce some relevant quantities; in the following spin-independent DM-nucleus interactions and the standard halo model are assumed.
The total DM recoil rate $R$ is given by:
\begin{equation}
\label{eq:tot_rate}
R= \sum_J \int_\Omega d\Omega' \int_{E_{\text{thresh}}}^{E_{r\text{,max}}^J} dE_r \ \frac{\partial^2 {\cal R}^J}{\partial\Omega' \partial E_r} \,,
\end{equation}

where $J$-index refers to the target atom species and $\frac{\partial^2 {\cal R}^J}{\partial\Omega' \partial E_r}$ is the double-differential rate with respect to the solid angle and the recoil energy. The recoil energy integration range spans from zero (or the cutoff induced by the experimental sensitivity $E_{\text{thresh}}$) to $E_{r\text{,max}}^J$, which is the maximum recoil energy per atomic species:
\begin{equation}
E_{r\text{,max}}^J= \gamma^J\, \frac{1}{2} M_{\chi} v_{\text{max}}^2\ , \qquad \gamma^J= \frac{4\mu_{\chi J}^2}{M_{\chi}\, M_{J}} \,.
\end{equation}
Here $v_{\text{max}}$ is the maximum possible DM velocity and is set to the sum of the galaxy escape velocity $v_{\text{esc}}$ and the Sun's velocity in the galactic rest frame $v_0$; $M_{\chi}$ and $M_{J}$ represent the DM and the $J$-th atomic species masses respectively, and $\mu_{\chi J}$ is the corresponding DM-nucleus reduced mass.\\
To predict the rate of events producing a defect, we must consider that the scattered atom is forced to generate a Frenkel pair when the recoil energy is above the TDE surface. Thus, the so-called defect rate $R_{\text{Def}}$ has the same expression as in Eq. \eqref{eq:tot_rate}, but a different energy integration range:
\begin{equation}
\label{eq:defect_rate}
R_{\text{Def}} = \sum_J \int_\Omega d\Omega'\, \int_{\text{TDE}^J(\Omega')}^{E_{r\text{,max}}^J} dE_r \, \frac{\partial^2 {\cal R}^J}{\partial\Omega' \partial E_r} \, ,
\end{equation}
where TDE$^J(\Omega')$ represents the TDE for the $J$-th species as the PKA at the solid angle position $\Omega'$. The single Na and I contributions can be separated by considering that $R_{\text{Def}}= \sum_J R_{\text{Def}}^{J}$.\\
Further details regarding the evaluation of the DM scattering rate are reported in Appendix \hyperref[AppendixB]{B}.
\begin{figure}
    \centering
    \includegraphics[width=\columnwidth]{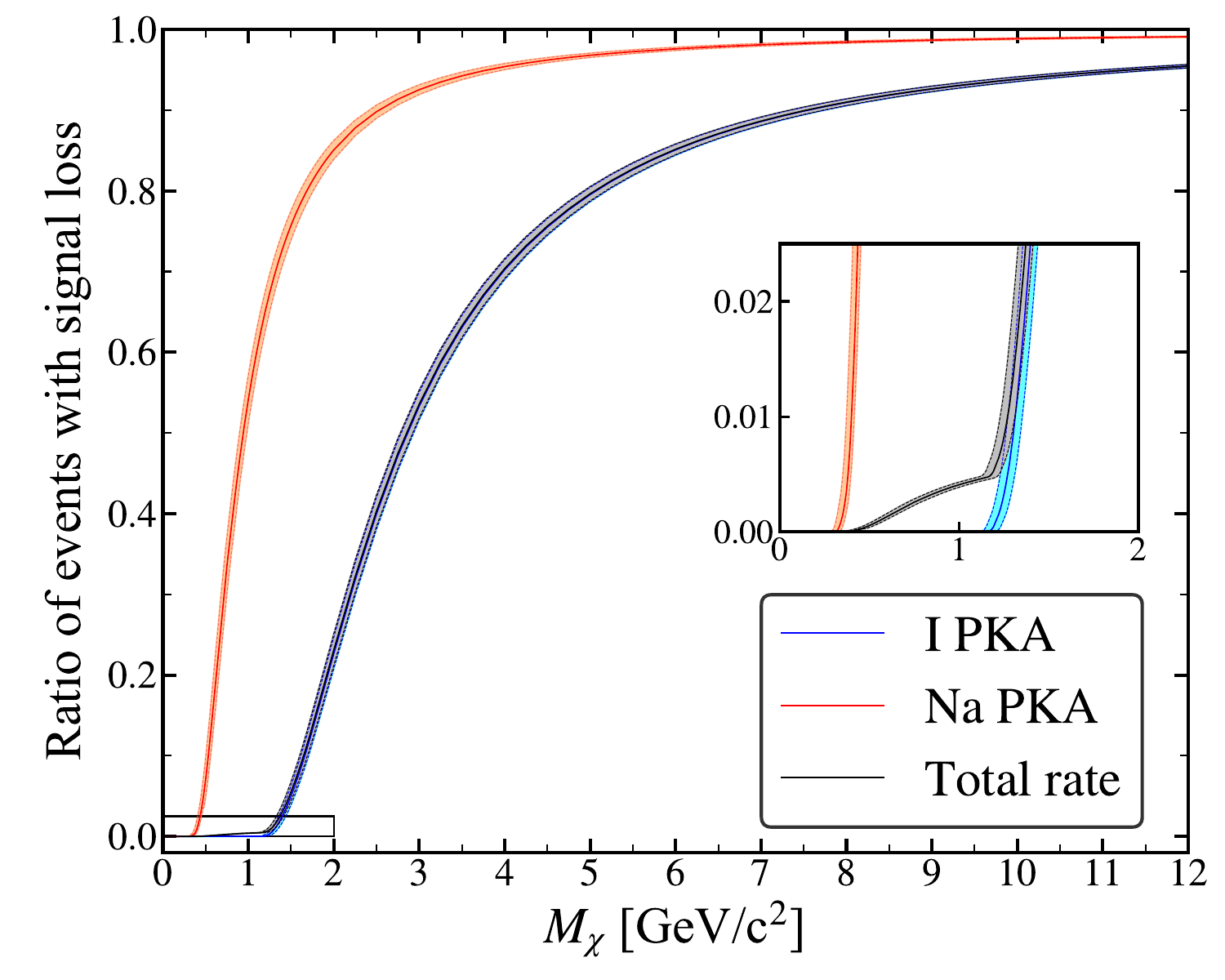}
    \caption{Ratio of events with a signal loss due to the formation of Frenkel pairs, as a function of the DM mass and depending on the PKA. The red (blue) line shows the result obtained using the calculated TDE for Na (I), while the dashed lines correspond to a $\pm$5~eV uncertainty on the corresponding TDE.
    The black line represents the ratio between the total rates $R_{\text{Def}}/R$. The inset shows a zoom of the plot at lower masses.}
    \label{fig:EventRatio}
\end{figure}
\begin{figure}
    \centering
    \includegraphics[width=\columnwidth]{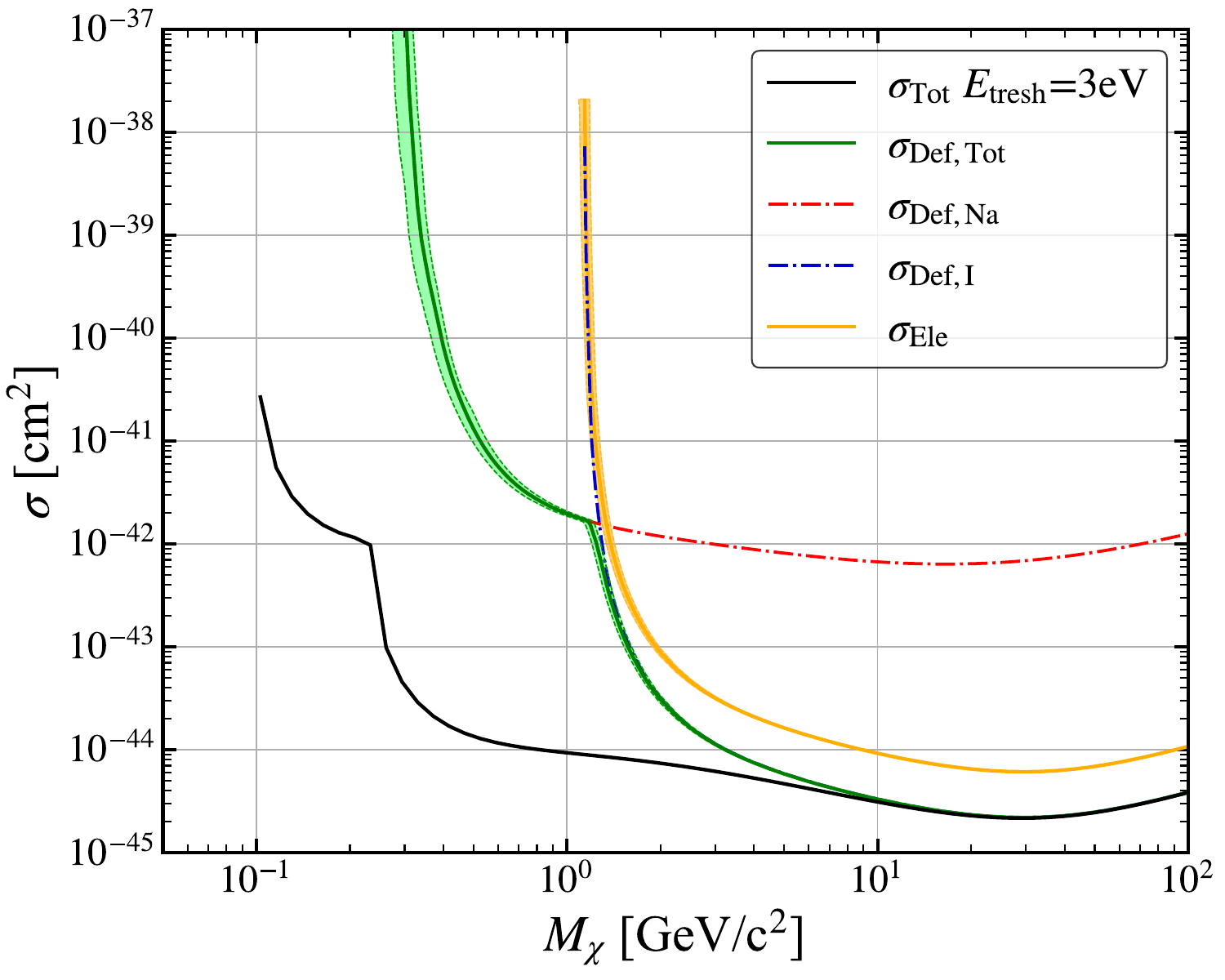}
    \caption{Experimental reach as a function of the DM particle mass, assuming a 1~kg$\cdot$yr exposure, and a background-free experiment. In black the total cross section is plotted assuming a sensitivity threshold of 3~eV; in green the defects cross section for both Na and I is shown, the shaded region corresponds to the uncertainty window on the TDE estimate. In blue and red, dash-dotted lines, the defects cross section for I and Na individually, respectively. In yellow-orange the lower estimate for the electronic channel cross section is reported.}
    \label{fig:Crosssection}
\end{figure}

In Fig.~\ref{fig:EventRatio} we report the ratio of DM events forming at least a Frenkel pair, as a function of the DM mass, which is given by the ratio $R_{\text{Def}}/R$. The same ratios, evaluated for the two PKA atoms, are also shown.\\
The starting DM mass that enables the defects production is around 0.4~GeV/c$^{2}$ for Na PKA, while for I PKA it is 1.5~GeV/c$^{2}$.
The ratio of events generating at least one defect increases for higher DM masses, reaching a value of 90\% at around 2.5~GeV/c$^{2}$ in the Na case, and 7.5~GeV/c$^{2}$ for I. We can also easily understand the difference in sodium and iodine behavior. Iodine recoil energies are smaller, and its TDE is higher; thus at the same DM masses the relative number of events forming defects will be smaller too.  
Due to the $A_{J}^2$ enhancement factor in the differential rate, the iodine contribution dominates, making the total defect rate closely aligned with the iodine contribution, particularly for larger DM masses. Similarly, the ratio $R_{\text{Def}}/R$ largely reflects the iodine contribution, with only a minor deviation at lower masses, as shown in the inset of Fig.~\ref{fig:EventRatio}.\\
The shaded bands in Fig.~\ref{fig:EventRatio} represent the variation in the rates when the TDE is varied within a $\pm$5 eV uncertainty. This range corresponds to the maximum energy interval over which defect formation was explicitly checked. As shown, this uncertainty produces only negligible changes in the rates, demonstrating the robustness of our results.
\indent We now move to the prediction of the experimental reach of an NaI-based detector searching for DM-nuclei scattering events. We assume that the detector is capable of identifying all events leading to defect formation through one of the detection mechanisms previously described.
The result obtained assuming an exposure $\mathcal{E}=$ 1~kg$\cdot$yr and a background-free experiment is shown in Fig.~\ref{fig:Crosssection}. 
Given the background-free assumption, imposing a confidence level (C.L.) of 95\% corresponds to the detection of 3 events, thus the cross section limits have been set using the expression:
\begin{equation}
    \sigma = \frac{3}{(R/\sigma_{\text{SI}}) \mathcal{E}} \ .
\end{equation}
The black curve in Fig.~\ref{fig:Crosssection} represents the experimental reach for an experiment with a sensitivity threshold of 3~eV, in which all the DM-nuclei scattering are detectable, and is computed using Eq. \eqref{eq:tot_rate} rate. In green, we report the reach for an experiment based on the detection of DM events that generate at least one defect, of whatever type, using their effects on the different detection channels. The rate from Eq. \eqref{eq:defect_rate} is thus adopted. The reach obtained considering only the Na or I defect rates is shown in red and blue, respectively. The DM masses at which the rate of defect-forming events becomes significant are clearly visible, as is the shift in the main contributor from Na to I atoms.\\
In yellow-orange we present an estimate of the experimental reach for the electronic channel, which relies on the previously discussed "elevator state" or photodetector mechanisms, enabled by the Iodine defect in-gap states.\\
This estimate has been performed by considering only the Iodine PKA recoils with direction in the solid angle fraction that leads to interstitial I defects, which we refer to as $\Omega_{\text{I}}$. Then we can approximate the rate of events forming in-gap states as:
\begin{equation}
    R_{\text{Ele}}\sim R_{\text{Def}}^{\text{I}}\Big(\frac{\Omega_{\text{I}}}{4\pi}\Big) \ .
\end{equation} 
In Fig.~\ref{fig:Crosssection} the result for $\Omega_{\text{I}}=0.36\cdot(4\pi)$ is shown.
This is a lower estimate for the events rate, since it does not take into account events at higher released energy that generate multiple defects, with at least one being an interstitial I atom, for directions and PKAs that at lower energies only generate an interstitial Na atom. The correct rate would also have to consider these contributions, which cannot be easily estimated, and would lower the possible detectable cross section, matching for large DM masses the total cross section shown as a black line in Fig.~\ref{fig:Crosssection}.

\section{Conclusions}
In this work, the formation of defects in NaI crystals induced by dark matter collisions and their implications for direct detection experiments have been investigated. By combining molecular dynamics simulations and density functional theory calculations, we characterized the main defects properties, including their formation energies, the anisotropic threshold displacement energy and their effect on the system's electronic states. In particular, we observe a modification of the states close to the valence band maximum of NaI induced by all the analyzed defects; notably, in the case of an iodine Frenkel pair, new electronic states emerge within the band gap.
These states allow transitions at energies lower than the band gap, opening  potential novel detection channels in NaI-based cryogenic scintillating calorimeters, that can enhance their sensitivity to dark matter particles in specific mass ranges.\\
Furthermore, we have computed the DM-nuclei scattering rates and calculated the ratio of events that will produce a signal loss in the phonon channel due to defect formation, which can be also enhanced by the presence of localized vibrational modes.\\
Our study highlights the need for further investigations into defect formation and defect-induced charge signals in NaI-based detectors. Understanding their impact on the phonon and scintillation light channels is crucial, particularly in relation to aging effects during long-term experiments.

\section*{Data availability statement}
The data that support the findings of this study are available upon reasonable request from the authors.

\section*{Acknowledgments}
The \mbox{COSINUS} collaboration/authors thanks the Director and the Computing and
Network Service of the Laboratori Nazionali del Gran Sasso (\mbox{LNGS-INFN}).
This research used resources of the LNGS HPC cluster realised in the
framework of Spoke 0 and Spoke 5 of the ICSC project - Centro Nazionale di
Ricerca in High Performance Computing, Big Data and Quantum Computing,
funded by the NextGenerationEU European initiative through the Italian
Ministry of University and Research, PNRR Mission 4, Component 2:
Investment 1.4, Project code CN00000013 - CUP I53C21000340006.

\section*{APPENDIX A: N\MakeLowercase{a} defect and localized modes}\label{AppendixA}
\begin{figure}
    \centering
    \includegraphics[width=\columnwidth]{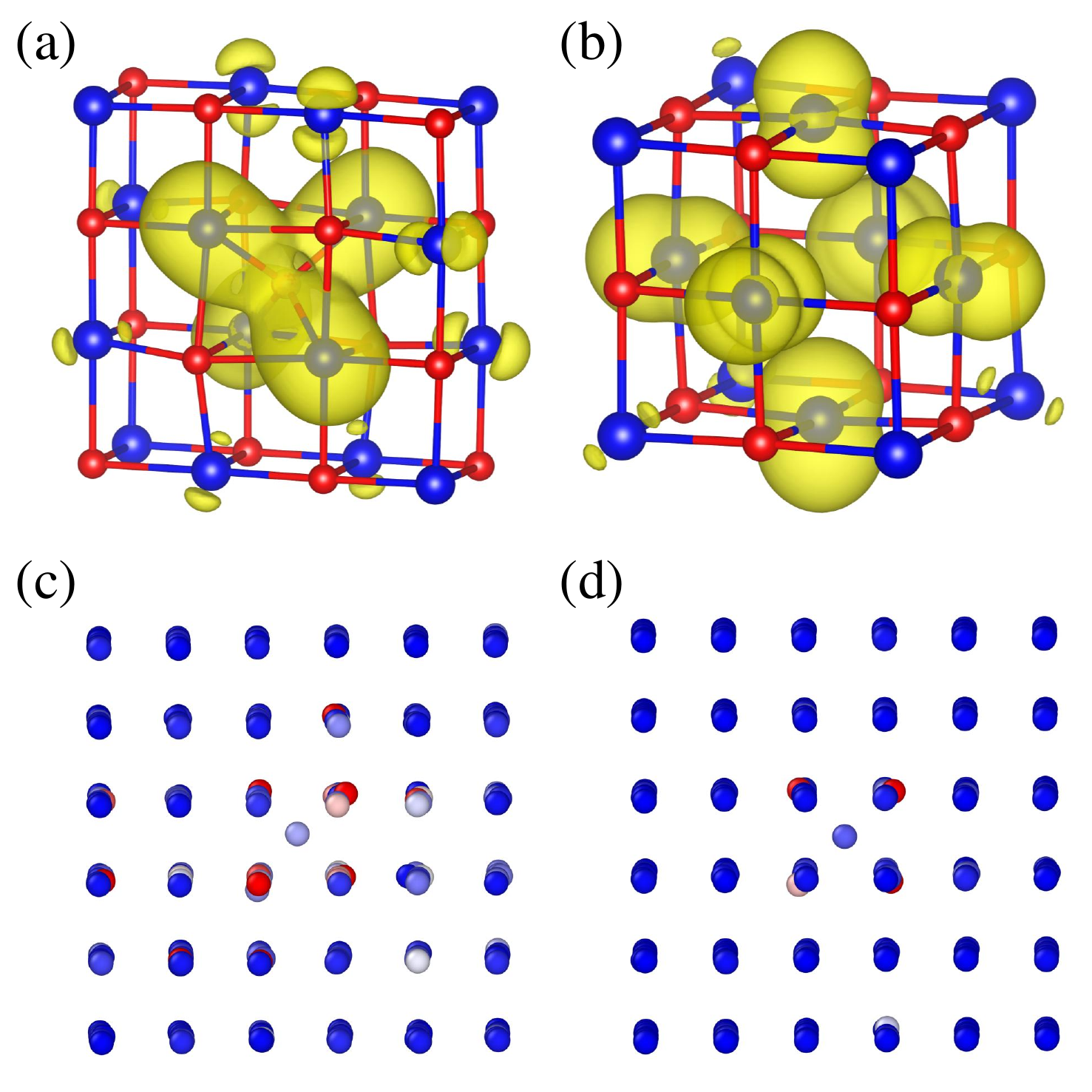}
    \caption{(a-b) Sodium interstitial atom and its corresponding vacancy. Na atoms are shown in red while I atoms are in blue. The yellow surfaces indicate the charge density associated to the states present in the energy range [-3.4,-2.8]~eV for panel (a) and [-0.6,0.0]~eV for panel (b), relative to the corresponding Fermi energy (red dashed line in Fig.~\ref{fig:defect-dos} (c)). (c-d) Snapshots of the molecular dynamics simulation of a 26~eV DM collision along the [110] direction on an Na PKA. Panel (c) is taken after 1~ps from the collision, while panel (d) after 20~ps. The color coding refers to the atom's kinetic energy, going from blue (lowest energy) to red (highest energy).}
    \label{fig:Nacharge}
\end{figure}

The electronic configuration associated with an Na defect does not introduce new states in the band gap, as shown in Fig.~\ref{fig:defect-dos}~(c). However, significant changes in the electronic states are still observed.\\
By analyzing the projected density of states, two energy regions were identified where the electronic states are predominantly localized around the atoms near the interstitial Na atom and the vacancy.  In Fig.~\ref{fig:Nacharge}~(a-b) the real space charge distributions corresponding to the energy intervals [-3.4,-2.8]~eV (panel~(a)) and [-0.6,0.0]~eV (panel~(b)) are shown. These distributions clearly show localization around the interstitial Na atom and the vacancy, respectively. The states around the vacancy are particularly interesting since they influence the DOS near the Fermi energy, potentially altering transport properties and scintillation processes.\\
While DFT calculations are needed to obtain information on the electronic states, MD simulations offer a valuable advantage in tracking the system's time evolution, even after the formation of the Frenkel pair. By extending the simulation time range, we observed the presence of localized vibrational modes centered on the interstitial atom. Wavelet analysis of these vibrations revealed an associated frequency above the equilibrium NaI phonon energy range. 
In Fig.~\ref{fig:Nacharge}~(c-d) two MD snapshots at different times are shown, with atoms colored according to their kinetic energy. The deposited energy is 26~eV on an Na PKA thus resulting in the formation of an Na Frenkel pair, with the interstitial atom visible in both panels. Panel~(c) corresponds to 1~ps after the collision, while panel~(d) shows the state after 20~ps. While the majority of atoms lose their kinetic energy through thermalization processes, the atoms near the interstitial Na still keep vibrating after 20~ps.
This anomalous behavior persists even after hundreds of ps and can be naturally explained considering the formation of localized vibrational modes with characteristic frequencies lying outside the phonon energy range and weakly coupled with the host phonon modes.\\
All these properties closely resemble the features induced by intrinsic localized modes (ILMs), also called discrete breathers, for which NaI crystals have been identified in the literature as a prime candidate to host them~\cite{ILM_1,ILM_2,ILM_3,ILM_4,ILM_5}.
The possibility of long-lasting intrinsically localized modes in the NaI detector could lead to a slower reset of the experimental signal peak. Additionally, these modes may introduce a thermal component that remains significant on the millisecond timescale, particularly in higher energy events involving the formation of multiple defects.

\section*{APPENDIX B: DM scattering rate evaluation}\label{AppendixB}
We consider elastic nuclear recoils with non-relativistic DM particles; the interaction is assumed to be spin-independent by imposing a neutron/proton symmetry of the couplings ($f_n \equiv f_p$).
The interaction rate can then be written in terms of the standard spin-independent cross section $\sigma_{\text{SI}}$:
\begin{equation}
\sigma_{\text{SI}}=\frac{f_n^2  \mu_{\chi n}^2}{\pi} \,, 
\end{equation}  
where the $\mu_{\chi n}$ parameter stands for the DM-nucleon reduced mass. In our work we have set $\sigma_{\text{SI}}=10^{-39}~\text{cm}^{2}$.\\
The differential rate expresses the number of events per unit of target mass, time, solid angle and recoil energy, and is given by ~\cite{diffratecit,dmratecit,emken2019,LEWIN199687}:
\begin{equation}
\begin{split}
\frac{\partial^2  {\cal R}}{\partial \Omega \partial E_r} &= \frac{\rho_0\, \sigma_{\text{SI}}}{4 \pi\, M_\chi \, \mu_{\chi n}^{2}} \sum_{J} X_J \, A_J^2  \,  F_J (E_r,\,A_J)^2\, \\ 
&  \times \hat{f}_{\text{lab}} (v_{\text{min}}, \mathbf{\hat{q}}, t)\,,
\end{split}
\end{equation}
where $ M_\chi$ is the DM mass, $\rho_0$ is the DM energy density fixed to $0.3$~GeV~cm$^{-3}$, the $J$-index stands for the atomic species in the target material, $X_J$ is the mass ratio of the $J$-th atomic species $X_J= \frac{M_J}{\sum_{J} M_J}$, $v_{\text{min}}=\sqrt{2M_{J}E_{r}}/2\mu_{\chi J}$ is the minimum DM speed needed to produce a recoil of energy $E_{r}$ on a nuclei of the $J$-th species.
We fixed the nuclear form factor $F_{J}$ as the nuclear Helm form factor for the $J$-th species~\cite{HelmFormcit}:
\begin{equation}
F_J (E_r,\, A_J)=  \frac{3}{k \,{\cal Z}} \;j_1 \left(k \,{\cal Z}\right) \, e^{-\frac{1}{2}(k\, s)^2}\,,
\end{equation}
where $q= \hbar \, k$ is the transferred momentum $\sqrt{2 M_J E_r}$, $j_1$ indicates a Bessel function of the first kind and $\cal Z$ is the effective nuclear radius:
\begin{equation}
\begin{split}
& {\cal Z}= \sqrt{c^2 +\frac{7}{3} \pi^2 a^2- 5\, s^2} \, ,\\
& c= \big(1.23 A_{J}^\frac{1}{3}- 0.6\big) \; \text{fm} \, ,\\
& s\approx 0.9 \; \text{fm} \, ,  \ \ \ a\approx 0.52 \; \text{fm} \, . 
\end{split} 
\end{equation}

We point out that the nuclear form factor is not crucial in determining the ratio of recoils producing defects in the crystal, but has an effect on the experimental reach for larger DM masses.\\
Finally $\hat{f}_{\text{lab}}$ is the Radon transform along $\hat{\textbf{q}}$ of the Maxwell distribution describing the DM velocity density function~\cite{Halocit,dmratecit}:
\begin{equation}
\hat{f}_{\text{lab}}= \frac{1}{N_\text{esc} \sqrt{2\pi\sigma_v^2}} \left[ e^{-\frac{|v_{\text{min}}+\hat{\textbf{q}}\cdot \textbf{v}_{\text{lab}}|^2}{2\sigma_v^2}}-e^{-\frac{v_{\text{esc}}^2}{2 \sigma_v^2}} \right],
\end{equation}
\\
where $\hat{\textbf{q}}$ is the recoil direction in the detector coordinates, $\textbf{v}_{\text{lab}}$ is the velocity of the laboratory relative to a stationary observer, $v_{\text{esc}}$ = 544~km/s is the circular escape velocity at the Solar System's distance from the Milky Way's center, $\sigma_v = \frac{v_{0}}{\sqrt{2}}$ is the DM velocity standard deviation, $v_{0} =$ 260~km/s is the Milky Way's circular speed, and $N_{\text{esc}}$ is a normalization factor. Finally, the scalar product $\hat{q} \cdot \textbf{v}_{\text{lab}}$ gives all the information regarding the Earth's motion relative to the DM wind, and thus the daily and annual modulation of the events rate.

\bibliography{Bibliography}

\end{document}